\documentclass[12pt,preprint,english]{aastex}
\usepackage[utf8]{inputenc}
\usepackage[T1]{fontenc}
\usepackage{amsmath,amssymb,mathrsfs}
\usepackage{babel}
\usepackage{graphicx}
\usepackage{rotating}
\usepackage{caption2}
\newcommand{\sppr}{\ensuremath{\textrm{MSA}}}
\newcommand{\bigp}[1]{\ensuremath{\left(#1\right)}}

\newcommand{\bigpf}[1]{\ensuremath{\left\{#1\right\}}}
\newcommand{\dd}[2]{\ensuremath{\frac{\mathrm{d}#1}{\mathrm{d}#2}}}
\renewcommand{\deg}{^{\circ} }
\renewcommand{\vec}{\mathbf}

\title{Using an Assumption about the Monotony of Spiral Arms to Determine the Orientation Angles of Galaxies}
\author{S.G.Poltorak and A.M. Fridman}
\affil{Institute of Astronomy of the Russian Academy of Science, \\
Pyatnitskaya ul. 48, Moscow, 109017 Russia \\
\email{poltorak@alsenet.com}
}

\begin{document}

\begin{abstract}
A method is proposed for the determination of the position and inclination angles of the plane of a spiral galaxy based on the assumption that every  spiral arm is a monotonic function of the radius versus azimuthal angle. This method may yield more accurate results than the more commonly employed isophote method, which is fraught with various drawbacks. The use of the new method is illustrated by applying it to a sample of 43 objects, and the results agree well with data from other sources.
\end{abstract}

\maketitle


\section{INTRODUCTION}
The orientation of a galactic disk with respect to the line-of-sight is one of the most important parameters of a spiral galaxy required for studies of its morphology and dynamics. This orientation is determined by the position angle $PA$ and inclination angle $i$ of the plane of the disk relative to the plane of the sky. Several methods for determining these angles are known, some making use of kinematic data and other based on photometry. The latter category includes the most widely used method for determining the inclination angle $i$ of a galactic plane relative to the plane of the sky~--- the so-called isophote method. This method estimates $i$ from the degree of elongation of the observed elliptical galaxy isophotes, with or without allowance for the finite thickness of the disk. However, the results obtained with this method have large uncertainties if the brightness distribution in the galaxy deviates strongly from axial symmetry. In fact, this is the case in many observed galaxies, due to the presence of bright spirals, a bar, and/or dust lanes; for isophotes in regions of the disk at large galactocentric distances, warping of the disk can also contribute to deviations from axial symmetry. The method that we propose here can yield more accurate results in such cases.

All the methods employed assume that the galactic disk can be described using some model. Our method for determining the angles $PA$ and $i$ is based on a simple and very general assumption about the shape of every spiral arm. We assume the spiral arm to be monotonic; i.e., the radius vector $\vec R$ of the spiral in the galactic plane increases monotonically with $\theta$: $\dd{R}{\theta}>0$. Section \ref{The Idea} describes the main features of the method and justifies the adopted assumptions. Section \ref{Testing} describes analytically the capabilities of the method as applied to logarithmic spirals, which are simplest in terms of the mathematics involved. Section \ref{Comparison} compares the method with other methods and discusses its advantages and disadvantages, while Section \ref{Application to Ohio sample} presents the results of applying the new method to a sample of 43 galaxes from the Ohio State University Bright Spiral Galaxy Survey. We present the derivation of all formulae used in the Appendix.

\section{IDEA BEHIND THE METHOD}\label{The Idea}
The method we describe here is based on the intuitively clear assumption that all spiral arms except for ring-shaped structures are monotonic. In other words, the radius increases monotonically in the course of moving along a spiral from the center of the galaxy toward its periphery. Note that this should be true not only for ideal, two-armed (or many-armed) spirals, but also for branching spirals. Spirals can be traced using young stars (in $\textit{B}$ images), gas ($\textrm{H}\alpha$ and \textrm{NII} images), or even dust lanes ($\textit{B}$, $\textit{V}$, and $\textit{R}$ images). In this case, the old stellar population is of little use due to the relatively large width of the corresponding spiral arms.

\begin{figure*}[htb]
  \begin{center}
    \begin{picture}(0,0)
     \put(50,133){\begin{turn}{19}Galaxy plane\end{turn}}
     \put(225,137){\begin{turn}{18}Sky plane\end{turn}}
     \put(265,117){$r$}
     \put(270,100){$\varphi$}
     \put(167,30){$i$}
     \put(135,127){$R$}
     \put(170,105){$\theta$}
     \put(315,78){\begin{turn}{-5}Line-of-sight\end{turn}}
    \end{picture}
    \includegraphics[width=0.7\textwidth]{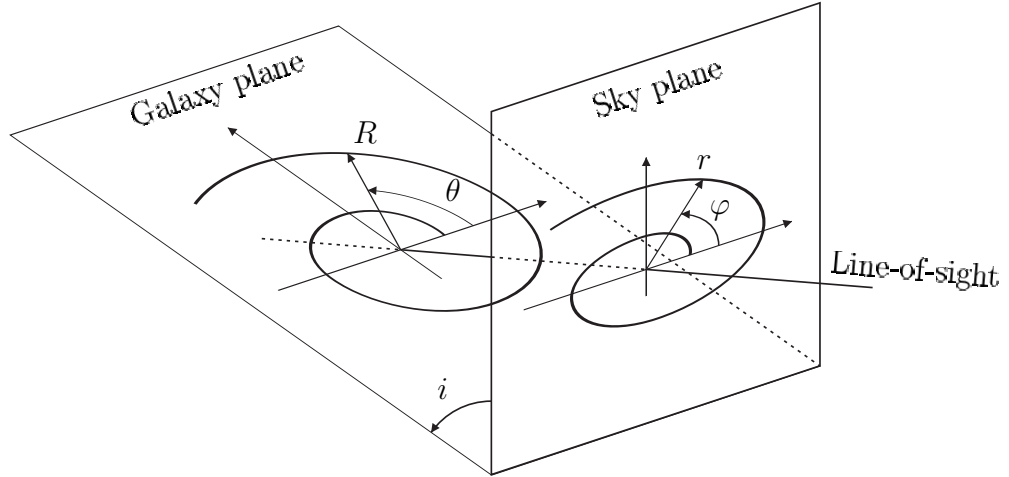}
    \caption{Schematic demonstrating how the monotonic behavior of the functions $R(\theta)$ and $r(\varphi)$ differs in the planes of the galaxy and of the sky. Whereas $R(\theta)$ is monotonic in the plane of the galaxy $\dd{R}{\theta} > 0$ , $r(\varphi)$ may be non-monotonic; i.e., $\dd{r}{\varphi}$ may change its sign. This is especially obvious in the case of small pitch angles and large inclination angles $i$. Note that we use the polar coordinates $R, \theta$ and $r, \varphi$ in the planes of the galaxy and of the sky, respectively.
    }
    \label{fig:galaxy-to-sky}
  \end{center}
\end{figure*}

The method proposed is not trivial. As is shown in Fig. \ref{fig:galaxy-to-sky}, when a spiral ia a monotonic function in the plane of the galaxy~---  i.e., the radius vector $\vec R$ of the spiral increases monotonically with the angle $\theta$~--- the same spiral may obey a non-monotonic function in the plane of the sky. Suppose that we choose the center of the galaxy as our coordinate origin, place several points with coordinates $r_i$ and $\varphi_i$ (the radius and azimuth in the plane of the sky) onto the spiral arm, and calculate for various $PA$ values ranging from $0\deg$ to $180\deg$ and $i$ values ranging from $0\deg$ to $90\deg$ corresponding coordinates $R_i(r_i, \varphi_i, PA, i)$ and $\theta_i(r_i, \varphi_i, PA, i)$ in the deprojected plane; the values of $PA$ and $i$ for which the spiral is monotonic (i.e., $\theta_i(R_i)$ is monotonic) then give the ranges of possible inclination and position angles for the given spiral.

\section{TESTING THE METHOD}
\label{Testing}
We will first analyze the capabilities of the method described, which we will refer to as the monotonic spiral arm (\sppr{}) method. As we pointed out above, this method can be applied to any spiral. However, to simplify the calculations, we analyze theoretically how it operates with a logarithmic spiral, $\theta=p\ln R$, where $p$ determines the pitch angle, defined as the angle between the tangent line to the spiral curve and a perpendicular to the radius vector at the given point. It is easy to show that, given the equation of the curve, the the tangent of the pitch angle is given by $\frac{1}{R}\dd{R}{\theta}=\frac{1}{p}$. It is obvious that the logarithmic spiral $\theta(R)$ is a monotonic function.

\begin{figure*}[ht]
  \begin{center}
    \begin{picture}(0,0)
     \put(130,-90){$i$}
     \put(25,-10){\begin{turn}{90}$\Delta PA_\sppr{}$\end{turn}}
     \put(360,-90){$i$}
     \put(255,-10){\begin{turn}{90}$\Delta i_\sppr{}$\end{turn}}
    \end{picture}
    \begin{tabular}{cc}
      \includegraphics[width=0.47\textwidth]{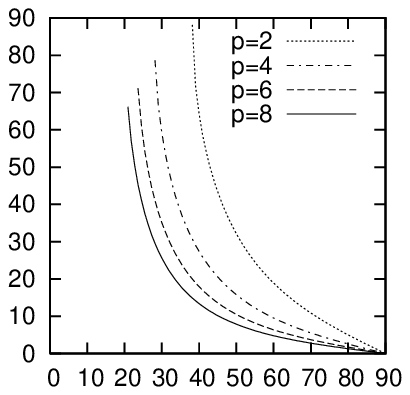} &
      \includegraphics[width=0.47\textwidth]{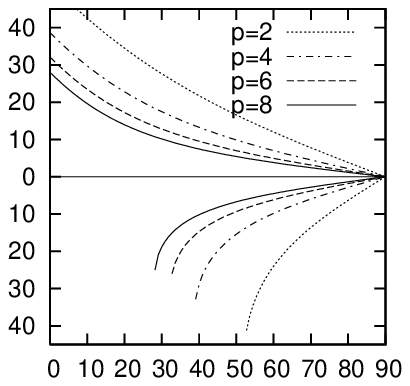}
    \end{tabular}
    \caption{Widths of the intervals of $PA$ (left) and $i$ (right) determined using the \sppr{} method for logarithmic spirals as functions of the true inclination angle $i$. The right plot shows two curves for each $p$: $\Delta i^{-}$ (the lower curve) and $\Delta i^{+}$ (the upper curve).
    }
    \label{fig:errors_analytical}
  \end{center}
\end{figure*}

Let us first note that a logarothmic spiral with pitch parameter $p$ inclined at an angle $i$ smaller than $i_\textrm{cr}=\arctan\frac{\sqrt{2(1+\sqrt{1+p^2})}}{p}$ appear monotonic even in the plane of the sky (see the Appendix \! \ref{i_critical_sec} for the derivation of this statement). Thus, for a face-on spiral (i.e., a spiral lying in the plane of the sky) the method described yields the range of inclination angles between $0$ to $i_\textrm{cr}$ . It can easily be shown that, for a spiral inclined by an arbitrary angle $i$, our method yields $PA_\sppr$ and $i_\sppr$ values in the intervals [see the Appendix for the derivation of \eqref{i_errors} and \eqref{PA_errors}]:
\begin{align}
	& \sqrt{1+\frac{2}{p^2}(1-\sqrt{p^2+1})}<\frac{\cos i_\sppr{}}{\cos i}<\sqrt{1+\frac{2}{p^2}(1+\sqrt{p^2+1})}, \label{i_area} \\
	& |\sin(PA_\sppr - PA)| < \frac{\cos i}{\sin^2 i}\sqrt{2\bigp{\frac{\sqrt{p^2+1}}{p}-1}}. \label{PA_area}
\end{align}
It is obvious that the true $i$ and $PA$ values are located within these intervals. Figure \ref{fig:errors_analytical} shows the widths of the resulting intervals of $i_\sppr{}$ and $PA_\sppr{}$ as functions of $i$ (it is clear that the $PA$ does not affect the result, since a change in $PA$ reduces to a simple turn of the entire system around the line-of-sight).

To apply the method described we wrote a code, which allows to select the center and indicate points onto spiral arms. The output is a drawing of angles found on $PA$--$i$ diagram. The program is available upon request (poltorak@alsenet.com).

We first applied our method to a simulated, but realistic galaxy image with logarithmic spiral arms. We adopted for the dependence of the brightness of
points on the galactic coordinates the relation
\begin{equation*}
	I(R, \theta) = e^{-R/R_0} + Ae^{-R/R_1}e^{-\bigp{\frac{\cos(\theta-p\ln R)}{\frac{\sigma}{R}+s R}}^2},
\end{equation*}
where $A, R_0, R_1, \sigma, s$, and $p$ are constants of the model\footnote{
	Here, we put meanings $A=0.7$, $R_0=R_1=100$, $\sigma=30$, $s=0.001$, and $p=2\div 7$. The image was $500\times500$ pixels in size and the galactic center was located at the center of the image. We included the term $\frac{\sigma}{R}$ to keep the width of the spiral arms approximately constant, while the term $s R$ slightly broadens the arms toward the periphery.
}.
We then inclined the simulated galaxy using different angles $PA$ and $i$.

Applying our method to these images demonstrated that:
\begin{itemize}
 \item[-] the method yields a domain of possible orientation angles for the simulated galaxy that contains the true $PA$ and $i$ values;
 \item[-] $\Delta i$ and $\Delta PA$ are in excellent agreement with the results of analytic computations [formulae \ref{i_area} and \ref{PA_area}];
 \item[-] the \sppr{} method is not very sensitive to the number of points on the arm (we recommend separating the points by at least 10 times the pixel size in order to reduce discretization errors);
 \item[-] to determine the angles $PA$ and $i$, it is sufficient to follow a half-turn of the spiral with six uniformly spaced points;
 \item[-] analyzing different parts of the same or several arms may improve the accuracy and indicate the presence of possible disk warps at the periphery (manifest as a change in $PA$ and $i$);
 \item[-] the uncertainty introduced by inaccurate choice of the center is usually significantly smaller than the derived widths of the angle intervals;
 \item[-] the \sppr{} methods works well for galaxies with large inclination angles and tightly wound spirals (i.e., with large $p$), but is essentially useless for almost face-on galaxies.
\end{itemize}

\section{COMPARISON WITH OTHER METHODS}\label{Comparison}
Having described the \sppr{} method, we now compare it with other methods for determining the orientation angles of galaxies.

\subsection*{Photometric Methods}
\begin{itemize}
\item{
\textbf{Fitting ellipses to galaxy isophotes.} \\
This method assumes that the galactic disk is axially symmetric, and therefore works well for early-type  galaxies or galaxies with low-contrast spirals. Since this method is sensitive to the presence of a bar or spiral structure (see, e.g., \cite{Stock-1955}), it can be applied only to the outer parts of galaxies, where many galaxies have warp and hence where this method is not very accurate.
}
\item{
\textbf{Fitting logarithmic or other forms of spirals to the observed spiral arms.} \\
This method can be applied to galaxies with well-developed spirals, i.e., late-type galaxies. However, logarithmic spirals do not describe all spirals well, especially, since the spiral arms in many galaxies branch, and often become open at the disk periphery. Like our method, this technique implicitly assumes that the spirals are monotonic, but imposes stricter limitations. The first version of this method was proposed by \cite{Danver-log-1942}.
}
\item{
\textbf{Maximizing the axisymmetric component of the Fourier expansion.} \\
This method is based on maximizing the zeroth or minimizing the second coefficient of the azimuthal Fourier expansion of the deprojected image introduced by the inclination; i.e., this method likewise assumes that the disk is axisymmetric. It also assumes that the second harmonic is only due to the disk inclination, which, naturally, is not true when the galaxy has a two-armed spiral pattern. However, as \cite{Barbera} showed for the BAG method described in their paper, this method is fairly reliable, even in the presence of a prominent spiral pattern.
}
\end{itemize}

\subsection*{Kinematic Methods}
\begin{itemize}
\item{
\textbf{Approximation of the velocity field using a model with purely circular rotation.} \\
The only widely used kinematic method for determining the orientation of a galaxy consists   
in fitting the two-dimensional observed velocity field with the velocity field of a model with
purely circular motions. However, \cite{Stock-1955}, \cite{Fridman-1997}, \cite{Lyakhovich-Vortex-eng}, and \cite{Fridman-15Gal} showed that the presence of a bar or spirals introduces systematic deviations into the derived $PA$ and $i$ values due to presence of non-circular motions. This method also cannot be applied to large galaxy samples, since it requires sophisticated measurements with a Fabry-Perot interferometer.
}
\end{itemize}

\subsection*{The Proposed Method}
The \sppr{} method can be classified as a photometric method. However, unlike other methods, it is based on a visual processing of the image: the observer must choose points on the spiral arm. This approach has its disadvantages: an astronomer with an active imagination could obtain incorrect results by choosing the trace points to be in the regions of low signal-to-noise ratio. The advantages of the method include the fact that it does not require the image to be prepared for reduction (i.e., there is no need to mask foreground stars, subtract the sky background, etc.), that makes this method easy and fast to apply to large samples. As we mentioned above, our method is most accurate when applied to highly inclined galaxy disks with well defined tightly wound spiral structure. Note also that, unlike other methods (except fitting logarithmic spirals to the observed arms), our technique makes use of the shape of the spiral when determining the orientation angles.

\section{APPLICATION OF THE METHOD TO THE OHIO SAMPLE}\label{Application to Ohio sample}
\begin{figure*}[ht!]
  \begin{center}
    \begin{picture}(0,0)
     \put(125,10){$PA_\sppr{}$}
     \put(25,75){\begin{turn}{90}$PA_\textrm{Garc\'\i a-G\'omez et al.}$\end{turn}}
     \put(355,10){$i_\sppr{}$}
     \put(245,80){\begin{turn}{90}$i_\textrm{Garc\'\i a-G\'omez et al.}$\end{turn}}
     \put(125,-190){$PA_\sppr{}$}
     \put(25,-105){\begin{turn}{90}$PA_\textrm{HyperLeda}$\end{turn}}
     \put(355,-190){$i_\sppr{}$}
     \put(245,-100){\begin{turn}{90}$i_\textrm{HyperLeda}$\end{turn}}
    \end{picture}
    \begin{tabular}{cc}
      \includegraphics[width=0.47\textwidth]{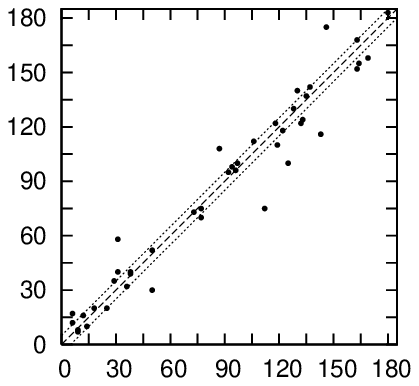} &
      \includegraphics[width=0.47\textwidth]{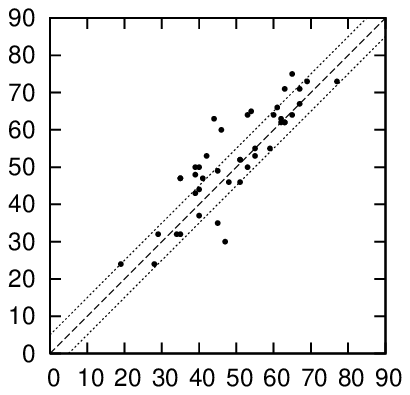} \\
      \includegraphics[width=0.47\textwidth]{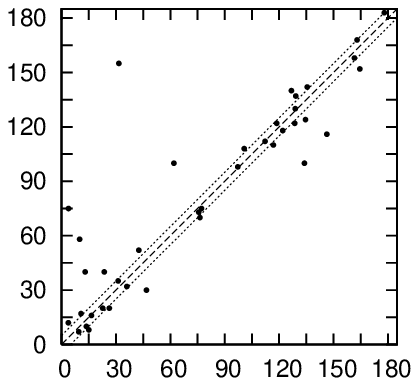} &
      \includegraphics[width=0.47\textwidth]{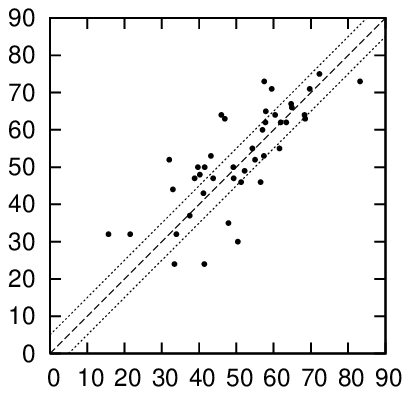} \\
    \end{tabular}
    \caption{
       Comparison of our calculated values of $PA$ (left plots) and $i$ (right plots) with those results of \cite{Garcia-Gomez} using their BAG method (upper plots) and those data from the HyperLeda catalog (lower plots). A $5\deg$ corridor is shown on both sides in all plots.
    }
    \label{fig:comparison_with_BAG_and_HyperLeda}
  \end{center}
\end{figure*}

\begin{figure*}[ht!]
  \begin{center}
    \begin{picture}(0,0)
     \put(120,-90){$i_\sppr{}$}
     \put(25,0){\begin{turn}{90}$\Delta PA_\sppr{}$\end{turn}}
     \put(350,-90){$i_\sppr{}$}
     \put(255,0){\begin{turn}{90}$\Delta i_\sppr{}$\end{turn}}
    \end{picture}
    \begin{tabular}{cc}
      \includegraphics[width=0.47\textwidth]{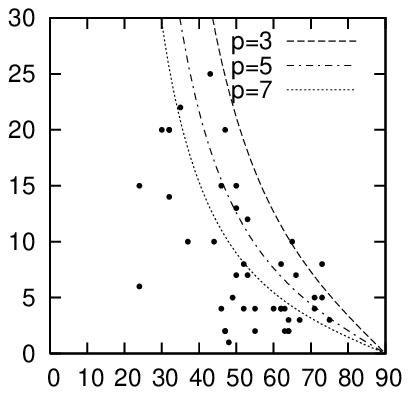} &
      \includegraphics[width=0.47\textwidth]{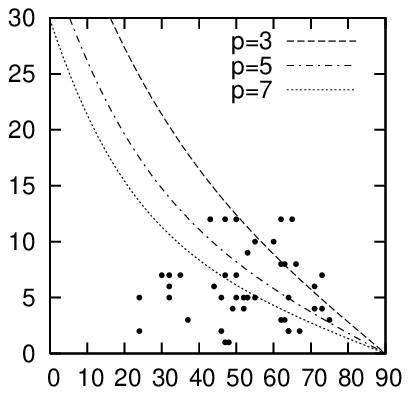} \\
    \end{tabular}
    \caption{
       Widths of the derived $PA_\sppr{}$ (left) and $i_\sppr{}$ (right) intervals as functions of $i_\sppr{}$ (dots). Also shown are the calculated widths for logarithmic spirals with various $p$ values. We consider the \sppr{} method to be inapplicable to galaxies with wide allowed angle intervals, and it is this that gives rise to the absence of points in the upper and right parts of the plots.
    }
    \label{fig:errors}
  \end{center}
\end{figure*}

To test our method on a large sample of real ojects, we applied it to the 43 galaxies from the State University Bright Spiral Galaxy Survey, and compared the results with those reported by \cite{Garcia-Gomez} and given in the HyperLeda catalog. Our results agree well with those of \cite{Garcia-Gomez}, obtained using the BAG method. The agreement with the HyperLeda data is slightly poorer, but this can be explained by the relatively modest quality of the data in this catalog (the data were adopted from various sources and are of various qualities; some of the data were adopted from photometric measurements in the RC3 catalog (\cite{RC3}), etc.). Figure \ref{fig:comparison_with_BAG_and_HyperLeda} compares our $PA$ and $i$ values with $PA_{BAG}$ and $PA_{HyperLeda}$, and $i_{BAG}$ and $i_{HyperLeda}$. Figure \ref{fig:errors} shows the errors obtained using our method as a function of the inclination angle $i$. As expected, the widths of the resulting domains depend on the inclination angle, but are smaller than those calculated in Section \ref{Testing}. This may be due to the small disk warps involved: we used various parts of all the spiral arms, and our method yielded the intersection of the domains for the individual spiral parts. This conclusion is supported by the fact that small (half-turn) parts of the spirals yield angle domains that are wider and somewhat different from those derived using the entire spiral.

\begin{figure}[htb]
  \begin{center}
    \includegraphics[width=0.65\columnwidth]{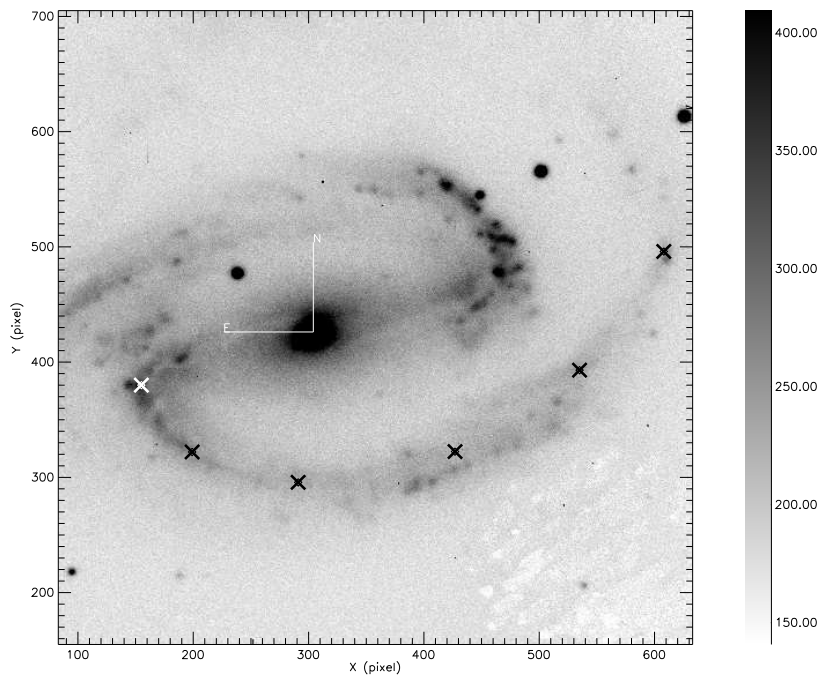}
    \caption{
           Application of the \sppr{} method to the Southern arm of NGC 1300: the B image with arm-tracing crosses (one is white to make it easier to see).
    }
    \label{fig:NGC1300-example-img}
  \end{center}
\end{figure}

\begin{figure*}[htb]
  \begin{center}
    \begin{picture}(0,0)
     \put(75,-85){$PA_\sppr{}$}
     \put(-5,0){\begin{turn}{90}$i_\sppr{}$\end{turn}}
     \put(260,-85){$PA_\sppr{}$}
     \put(185,0){\begin{turn}{90}$i_\sppr{}$\end{turn}}
    \end{picture}
    \begin{tabular}{ccc}
      \includegraphics[width=0.35\textwidth]{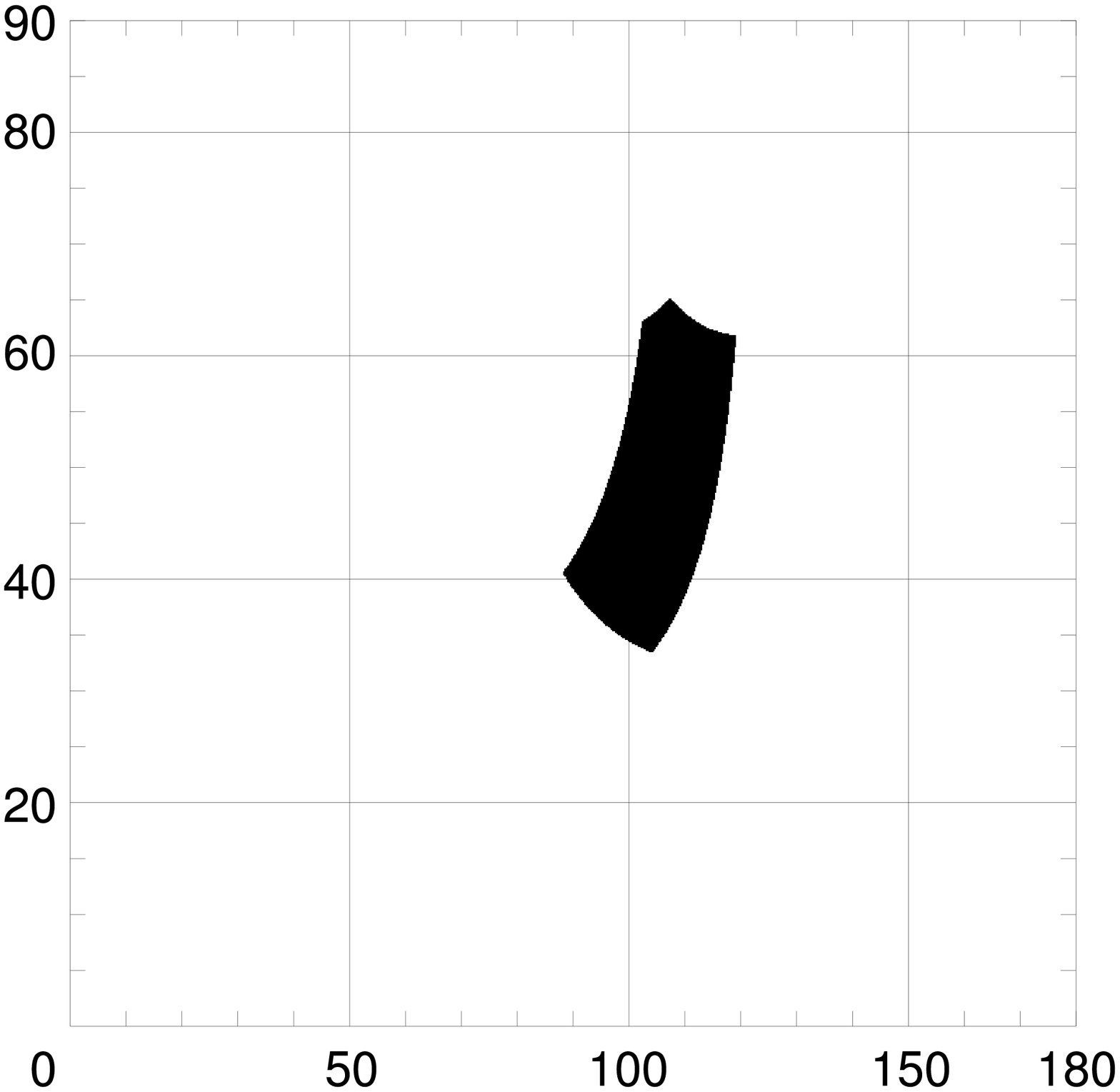} &
      \qquad &
      \includegraphics[width=0.35\textwidth]{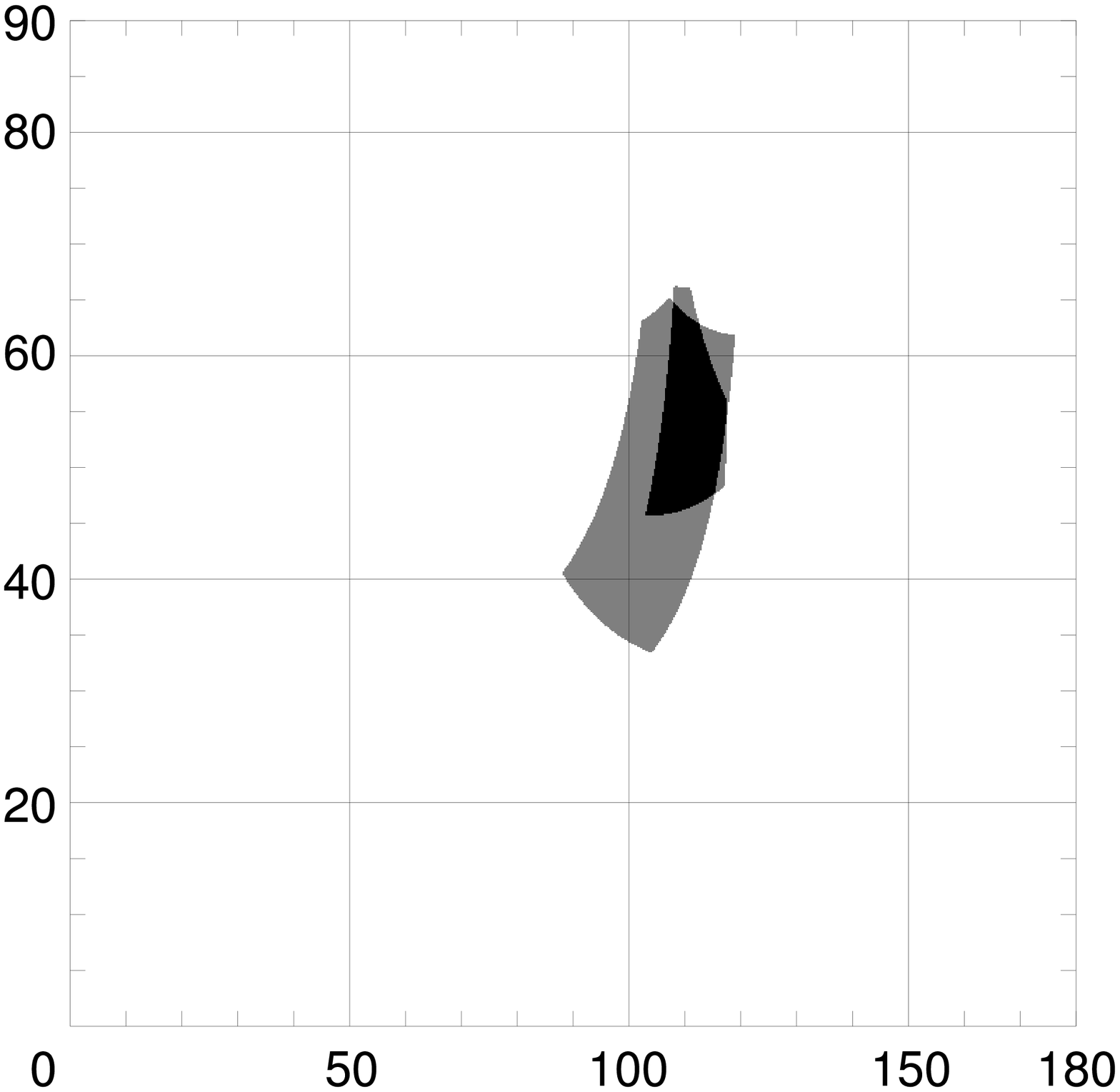} \\
    \end{tabular}
    \caption{
        $i$--$PA$ diagrams obtained by applying the \sppr{} method to NGC 1300. Left: result obtained by applying the method to the Southern arm only. Right: result obtained by applying the method to both the Northern and Southern arms. The black domain shows the intersection of the gray domains obtained for the individual arms. It is clear that both arms yield close values for $PA$ and $i$. The region of intersection makes it possible to reduce the intervals of possible values.
    }
    \label{fig:NGC1300-example-USP}
  \end{center}
\end{figure*}

As an example, Fig. \ref{fig:NGC1300-example-img} shows an image of the galaxy NGC 1300 with the tracing points superimposed, and Fig. \ref{fig:NGC1300-example-USP} shows the results of applying the method to this galaxy. A list of all galaxies whose orientation angles relative to the line-of-sight have been determined is presented in Table \ref{tbl:galaxies_orientations}.

\section{FURTHER DEVELOPMENT OF THE IDEA}
We propose to further develop this approach by adding an expansion of the image into a Fourier series
in angle. Note that a correct choice of the orientation angles would make the second harmonic of the deprojected image run exactly along the spiral arms. This would enable us to refine the angles $PA$ and $i$. It goes without saying that this proposed improved version of the method can work only for galaxies with symmetrical spirals. Automating this method is a fairly complicated task, but, in some special cases, the angles required to meet the criterion described can be chosen by hand.

\section{ACKNOWLEDGMENTS}
We thank V.L. Afanas’ev, A.A. Boyarchuk, S.N. Dodonov for fruitful discussions and A.V. Zasov for critical comments. This work was partially supported by the Russian Foundation for Basic Research (project code 05-02-17874-a), the Program of Support for Leading Scientific Schools of the Russian Federation (grant NSh. 7629.2006.2), and the Basic Research Program of the Department of Physical Sciences of the Russian Academy of Sciences \glqq Extended objects in the Universe\grqq. This research has made use of data of the Bright Spiral Galaxy Survey prepared
by researchers from Ohio State University, which is supported by the National Science Foundation
(grants AST-9217716 and AST-9617006) and Ohio State Universe.

\section{APPENDIX \\ DERIVATION OF FORMULAE}\label{Derivation of errors}
Here, we derive the formulas for the domains of possible orientation angles in the case of a logarithmic
spiral.

\subsection{Critical Inclination Angle}\label{i_critical_sec}
Let us now derive a formula for the critical inclination angle $i_\textrm{cr}$, such that, if tilted by no more than this angle a logarithmic spiral remains monotonic. We use two coordinate systems: polar coordinates $r$ and $\varphi$ in the plane of the sky and $R$ and $\theta$ in the plane of the galaxy. Let us now rotate the coordinate systems so that their $x$ axes coincide with each other and with the position angles (Fig. \ref{fig:galaxy-to-sky}).

We now consider a logarithmic spiral in the galactic plane:
\begin{align}\label{logarithmic-spiral}
	\theta = p\ln R + \theta_0 \textrm{ or } R = \exp\bigp{\frac{\theta-\theta_0}{p}}.
\end{align}
The coordinates $r$ and $\varphi$ can be written in terms of $R$ and $\theta$:
\begin{align}
	& r^2 = R^2(1-\sin^2\theta\,\sin^2 i), \label{r-via-R} \\
	& \tan\varphi = \tan\theta\,\cos i. \label{r-via-R-fi}
\end{align}
We now derive an equation for the projection of the logarithmic spiral onto the plane of the sky. We begin by using \eqref{r-via-R-fi} to write $\sin^2\theta$ as a function of the sky-plane coordinates:
\begin{align}\label{sin-theta}
	\sin^2\theta = \frac{\tan^2\varphi}{\cos^2 i + \tan^2\varphi}.
\end{align}
We now substitute \eqref{logarithmic-spiral} and \eqref{sin-theta} into \eqref{r-via-R} to obtain after
simple trigonometric manipulations
\begin{align}\label{r-via-varphi}
	r^2 = \frac{1+\tan^2\varphi}{1+\lambda^2\tan^2\varphi}\exp\bigp{\frac{2}{p}(\arctan(\lambda\tan\varphi)-\theta_0)},
\end{align}
where we introduced the quantity $\lambda=\frac{1}{\cos i}$.

Since $r > 0$ by definition, the condition that the spiral increase monotonically, $\dd{r}{\varphi}>0$, needed to determine $i_\textrm{cr}$ (or $\lambda_\textrm{cr}$) can be replaced by the condition
\begin{align}\label{dr2/dvarphi_r_ge_0}
	\dd{(r^2)}{\varphi} = 2r\dd{r}{\varphi} > 0.
\end{align}
We now introduce the derivative we need to derive:
\begin{align}\label{dr2/dvarphi}
	\dd{(r^2)}{\varphi} = & 2\frac{1+\tan^2\varphi}{(1+\lambda_\textrm{cr}^2\tan^2\varphi)^2}\bigpf{\tan^2\varphi\,\frac{1}{\lambda_\textrm{cr} p} + \tan\varphi\,\bigp{\frac{1}{\lambda_\textrm{cr}^2}-1} + \frac{1}{\lambda_\textrm{cr} p}}\times \nonumber \\
	& \times\exp\bigp{\frac{2}{p}(\arctan(\lambda_\textrm{cr}\tan\varphi)-\theta_0)}.
\end{align}

This quantity is greater than zero for any $\varphi$ if the discriminant of the quadratic equation in curly braces (an equation in $\tan\varphi$) is negative:
\begin{align}
	\mathscr{D} = \bigp{\frac{1}{\lambda_\textrm{cr}^2}-1}^2 - \frac{4}{\lambda_\textrm{cr}^2p^2} < 0 \textrm{\qquad or} \\
	p^2\lambda_\textrm{cr}^4 - 2(2+p^2)\lambda_\textrm{cr}^2 + p^2 < 0.
\end{align}
And we finally derive the condition for monotony of the projected, inclined logarithmic spiral:
\begin{align}\label{lambda-ineq}
	1+\frac{2}{p^2}(1-\sqrt{p^2+1}) < \lambda_\textrm{cr}^2 < 1+\frac{2}{p^2}(1+\sqrt{p^2+1}).
\end{align}

By definition,
\begin{align}
	\lambda_\textrm{cr}^2 = \frac{1}{\cos^2 i_\textrm{cr}} = 1 + \tan^2 i_\textrm{cr} \geqslant 1,
\end{align}
which means that the left-hand inequality in \eqref{lambda-ineq} is automatically satisfied for all $\lambda_\textrm{cr}$. Since $0\deg \leqslant i_\textrm{cr} \leqslant 90\deg$ we obtain
\begin{align}
	0 \leqslant \tan i_\textrm{cr} < \frac{\sqrt{2\bigp{1 + \sqrt{p^2 + 1}}}}{p}\label{i_critical}.
\end{align}
Hence, the inclinations of the galactic plane relative to the plane of the sky for which the observer will see the spiral as a monotonic function lie in the interval from $0$ to $\arctan\frac{\sqrt{2\bigp{1+\sqrt{p^2 + 1}}}}{p}$.

\subsection{Range of Inclination Angles}\label{i_errors_sec}
As for a face-on spiral, in the case of an arbitrarily inclined spiral, our method yields a range of possible inclination angles in the vicinity of the true value. To find this domain, let us consider a spiral inclined at an angle $i$ to the plane of the sky. We seek the range of inclination angles $i_\sppr$ such that deprojection yields a monotonic spiral. In addition to the coordinate systems used above, we introduce another coordinate system ($\tilde R, \tilde\theta$) in the plane onto which we deproject the galaxy using the angle $i_\sppr$:
\begin{align}\label{r-via-R_tilde}
	& r^2 = \tilde R^2(1-\sin^2\tilde\theta\,\sin^2 i_\sppr), \\
	& \tan\varphi = \tan\tilde\theta\,\cos i_\sppr.
\end{align}
We obtain, similarly to what we suggested in the preceding clause:
\begin{align}\label{R_tilde-via-theta_tilde}
	\tilde R^2 = \frac{1+\tan^2\tilde\theta}{1+\lambda^2\tan^2\tilde\theta}\exp\bigp{\frac{2}{p}(\arctan(\lambda\tan\tilde\theta)-\theta_0)},
\end{align}
where we have changed $\lambda$: $\lambda=\frac{\cos i_\sppr}{\cos i}$. In order for the deprojected spiral to be monotonic, it is necessary and sufficient that the derivative $\dd{(\tilde R^2)}{\tilde\theta}$ not change sign. Noting that \eqref{r-via-varphi} and \eqref{R_tilde-via-theta_tilde} are identical in form, we can immediately write the answer:
\begin{align}\label{i_errors}
	\sqrt{1 + \frac{2}{p^2}(1 - \sqrt{p^2 + 1})} < \frac{\cos i_\sppr}{\cos i} < \sqrt{1 + \frac{2}{p^2}(1 + \sqrt{p^2 + 1})}.
\end{align}
This is the final formula for the desired range of inclination angles yielded by our method.

\subsection{Range of Position Angles}\label{PA_errors_sec}
To find the range of position angles, we consider the spiral deprojected using the exact inclination angle $i$
and various position angles $PA_\sppr = PA + \Delta PA$.

Here, we introduce another supplementary coordinate system ($\bar R, \bar\theta$) in the plane rotated through the angles $i$ and $PA_\sppr$. We align the $x$ axis of this system along the assumed position axis, i.e., we rotate it by an angle $\Delta PA$ relative to the $x$ axis in the plane of the sky (which coincides with the direction of the true position axis) about the line-of-sight. The introduced coordinates can be related to the coordinates in the sky plane as follows:
\begin{align}\label{r-via-R_bar}
	& r^2 = \bar R^2(1-\sin^2\bar\theta\,\sin^2 i), \\
	& \tan\varphi = \frac{\cos\Delta PA\,\tan\bar\theta+\sin\Delta PA/\cos i}{-\sin\Delta PA\,\tan\bar\theta+\cos\Delta PA/\cos i}.
\end{align}
We similarly derive
\begin{align}\label{R_bar-via-theta_bar}
	\bar R^2 = f(\bar\theta)R^2 = f(\bar\theta)\exp\bigp{\frac{2}{p}(g(\bar\theta)-\theta_0)},
\end{align}
where
\begin{align}
	f(\bar\theta) = \frac{1+\tan^2\bar\theta}{1+\tan^2\bar\theta\,\cos^2 i + \sin^2 i\,(\tan\bar\theta\,\cos\Delta PA+\sin\Delta PA/\cos i)^2}, \nonumber \\
	\phantom{x} \nonumber \\
	g(\bar\theta) = \arctan\bigp{\frac{\tan(\varphi(\bar\theta))}{\cos i}} = \arctan\bigp{\frac{1}{\cos i}\frac{\tan\bar\theta\,\cos\Delta PA+\sin\Delta PA/\cos i}{-\tan\bar\theta\,\sin\Delta PA+\cos\Delta PA/\cos i}}. \nonumber
\end{align}
It can easily be shown that
\begin{align}
	\dd{g(\bar\theta)}{\bar\theta} = f(\bar\theta); \nonumber
\end{align}
it follows that
\begin{align}\label{dr2/dtheta_bar}
	\dd{(R^2(\bar\theta))}{\bar\theta} = \frac{2}{p}f(\bar\theta)R^2(\bar\theta).
\end{align}
The derivative of $f(\bar\theta)$ equals
\begin{align}
	\dd{f(\bar\theta)}{\bar\theta} = \frac{2f^2(\bar\theta)}{1+\tan^2\bar\theta}\frac{\sin\Delta PA\,\cos\Delta PA\,\sin^2 i}{\cos i}\bigp{\tan^2\bar\theta + \tan\bar\theta\,\tan\Delta PA\bigp{\cos i + \frac{1}{\cos i}} - 1}. \nonumber
\end{align}
As a result, we obtain
\begin{align}\label{dR_bar2/dtheta_bar}
	\dd{(\bar R^2)}{\bar\theta} = \dd{(R^2)}{\bar\theta}f(\bar\theta) + R^2 \dd{f(\bar\theta)}{\bar\theta} = \frac{\bar R^2}{f(\bar\theta)}\bigpf{\frac{2}{p}f^2(\bar\theta)+\dd{f(\bar\theta)}{\bar\theta}}.
\end{align}
This last quantity is greater than zero if and only if the expression in curly braces is positive:
\begin{align}\label{PA_errors-positive-condition}
	& \tan^2\bar\theta\bigp{\frac{\cos i}{p} + \sin^2 i\,\sin\Delta PA\,\cos\Delta PA} + \nonumber \\
	& + \tan\bar\theta\,\sin^2 i\,\sin^2\Delta PA\bigp{\cos i + \frac{1}{\cos i}} + \nonumber \\
	& + \bigp{\frac{\cos i}{p} - \sin^2 i\,\sin\Delta PA\,\cos\Delta PA} > 0.
\end{align}
The inequality \eqref{PA_errors-positive-condition} can be satisfied for any $\bar\theta$ value only if the discriminant of the quadratic equation (in $\tan\bar\theta$) is negative:
\begin{align}
	\mathscr{D} = \sin^4\Delta PA\,\sin^4 i\bigp{\cos i + \frac{1}{\cos i}}^2 + 4\bigp{\sin^2\Delta PA\,\cos^2\Delta PA\,\sin^4 i - \frac{\cos^2 i}{p^2}} < 0 \nonumber
\end{align}
or
\begin{align}\label{PA_errors}
	|\sin\Delta PA| < \frac{\cos i}{\sin^2 i}\sqrt{2\bigp{\frac{\sqrt{p^2+1}}{p}-1}}.
\end{align}
Hence, we find that, when applied to a logarithmic spiral with pitch parameter $p$ inclined at an angle $i$, our method yields the range of position-angle values near the true value from
$PA-\arcsin\bigp{\dfrac{\cos i}{\sin^2 i}\sqrt{2\bigp{\frac{\sqrt{p^2+1}}{p}-1}}}$ to $PA+\arcsin\bigp{\dfrac{\cos i}{\sin^2 i}\sqrt{2\bigp{\frac{\sqrt{p^2+1}}{p}-1}}}$.

\bibliography{gal_sp}

\begin{flushright}
 \textit{Translated by A. Dambis}
\end{flushright}

\begin{table}[hb!]
 \begin{center}
 \caption{The complete list of 43 galaxies whose $PA$ and $i$ have been determined using the \sppr{} method.} \label{tbl:galaxies_orientations}
 \vspace{12pt}
 \begin{tabular}{cc}
  \begin{tabular}{|l|r|r|r|r|}
	\hline
	Galaxy & $\phantom{\Delta}PA$ & $\Delta PA$ & $\phantom{\Delta}i$ & $\Delta i$ \\
	\hline
        NGC 150 & 110 & 7 & 66 & 8 \\
        NGC 157 & 40 & 4 & 60 & 10 \\
        NGC 210 & 168 & 8 & 52 & 5 \\
        NGC 289 & 122 & 2 & 47 & 1 \\
        NGC 488 & 8 & 1 & 48 & 1 \\
        NGC 613 & 122 & 2 & 63 & 3 \\
        NGC 685 & 100 & 20 & 47 & 12 \\
        NGC 908 & 75 & 4 & 62 & 8 \\
        NGC 1042 & 75 & 20 & 32 & 5 \\
        NGC 1073 & 155 & 15 & 24 & 5 \\
        NGC 1084 & 39 & 2 & 64 & 5 \\
        NGC 1187 & 137 & 13 & 50 & 12 \\
        NGC 1241 & 116 & 12 & 53 & 9 \\
        NGC 1300 & 108 & 2 & 47 & 7 \\
        NGC 1350 & 17 & 4 & 55 & 10 \\
        NGC 1371 & 124 & 5 & 49 & 4 \\
        NGC 1617 & 112 & 4 & 63 & 8 \\
        NGC 1637 & 40 & 10 & 44 & 6 \\
        NGC 1792 & 142 & 4 & 62 & 3 \\
        NGC 1964 & 35 & 5 & 71 & 6 \\
        NGC 2090 & 16 & 3 & 64 & 2 \\
        NGC 2196 & 30 & 25 & 43 & 12 \\
	\hline
  \end{tabular}
  &
  \begin{tabular}{|l|r|r|r|r|}
	\hline
	Galaxy & $\phantom{\Delta}PA$ & $\Delta PA$ & $\phantom{\Delta}i$ & $\Delta i$ \\
	\hline
        NGC 2280 & 152 & 3 & 67 & 2 \\
        NGC 2559 & 12 & 8 & 62 & 12 \\
        NGC 2964 & 98 & 10 & 65 & 12 \\
        NGC 3223 & 130 & 4 & 46 & 2 \\
        NGC 3261 & 70 & 10 & 37 & 3 \\
        NGC 3338 & 96 & 2 & 55 & 5 \\
        NGC 3423 & 58 & 20 & 32 & 7 \\
        NGC 3507 & 95 & 6 & 24 & 2 \\
        NGC 3511 & 73 & 8 & 73 & 7 \\
        NGC 3583 & 100 & 15 & 46 & 5 \\
        NGC 3596 & 175 & 14 & 32 & 6 \\
        NGC 3646 & 52 & 2 & 64 & 2 \\
        NGC 3675 & 183 & 4 & 71 & 4 \\
        NGC 3684 & 140 & 20 & 30 & 7 \\
        NGC 3686 & 20 & 15 & 50 & 5 \\
        NGC 3705 & 118 & 3 & 75 & 3 \\
        NGC 3726 & 10 & 7 & 50 & 7 \\
        NGC 3810 & 20 & 22 & 35 & 7 \\
        NGC 3877 & 32 & 5 & 73 & 4 \\
        NGC 3887 & 7 & 7 & 53 & 5 \\
        NGC 3893 & 158 & 4 & 52 & 4 \\
	& & & & \\
	\hline
  \end{tabular}
 \end{tabular}
 \end{center}
\end{table}

\end{document}